\documentclass[epj,twocolumn]{svjour}
\usepackage{amssymb}
\begin{document}
\title{On the robust thermodynamical structures against arbitrary entropy form and energy mean value}
\author{Takuya Yamano\thanks{\email{tyamano@o.cc.titech.ac.jp}}}
\institute{Faculty of Science, 
Tokyo Institute of Technology, 
Oh-okayama, Meguro-ku, Tokyo,152-8551, Japan}

\abstract{ We discuss that the thermodynamical Legendre transform structure can be retained not only for the arbitrary 
entropic form but also for the arbitrary form of the energy constraints by following the discussion of Plastino 
and Plastino. The thermodynamic relation between the expectation values 
and the conjugate Lagrange multipliers are seen to be universal. Furthermore, Gibbs' fundamental 
equation is shown to be unaffected by the choice of the entropy and the definition of the  mean values due to the robustness of the Legendre transform structure.    
}

\maketitle

\section{Introduction}
There exists the subtleness in understanding the relation between thermodynamics and 
statistical mechanics. Nowadays, on the other hand, some alternative entropic functionals to the conventional Boltzmann-Gibbs-Shannon\linebreak[4](BGS) entropy have been investigated intensively to apply to a variety of physical situations. 
Among them, as one example, Tsallis' information measure $S_q=(1-\int\!\! dxf^q)/(q-1)$\cite{88Tsallis,98Tsallis} where $q$ is a real parameter characterizing $S_q$ is attracting much 
attention. 
Thermostatistics based on this nonextesive measure has been shown to be useful for 
describing anomalous systems involving long-range interactions, long-term memory effect and 
(multi)fractal-like \linebreak[4]structure(see \cite{html} for concrete applications). 
Another example is Fisher's information measure $I=\int\!\! dx(f'^2/f)$
where $f$ is a normalized probability distribution\cite{Fisher}.
The connection between derivation of the variety of statistical laws of physics and the 
principle of minimum $I$ has been shown\cite{Fri}.
In view of these examples,
interests are increasing for exploring the possibility of the associated thermodynamics with 
non-BGS entropies. In this sense, therefore,
a certain mathematical relation among thermodynamical variables is considered to be in a 
crucial position in discussing the applicability of the alternative context. 
   
The Legendre transform structure(LTS)\cite{Callen,Beck,Schlogl} is considered to be a most fundamental 
relation which associates the phenomenologically based thermodynamics with the microscopically based statistical 
mechanics. Recently Plastino and Plastino\cite{APlast1,ARPlast} showed that the LTS is a
universal property independent of the selection of the entropic functional if only we adopt
the linear definition of an expectation value of observables. That is, weighting the 
each quantity with the probability corresponding to the configurational states preserves the 
LTS for an arbitrary form of the entropy functional. At the present stage, it should be of interest to investigate the structure against a nonlinear definition of the mean value in plurality of constraints.

In this paper, our discussion requires only the Jaynes Maximum entropy principle\cite{Jaynes1,Jaynes2,APlast2}. 
Jaynes' information theoretical approach to statistical mechanics based on the Shannon's 
extensive measure with a linear weighting of quantities as a mean value have successfully 
extended to Fisher's information measure(\cite{Frieden} and references therein). Moreover 
it is well known that Tsallis nonextensive measure with a nonlinear 
weighting has the LTS\cite{98Tsallis,Curado,95Tsallis1}. \linebreak[4]Therefore overall discussion based on the generic form of the entropy and a mean value with general weighting  
would illuminate the thermodynamical structures. It is the purpose of our present attempt to 
develop arguments along the line of the previous approach as mentioned above.

\section{The thermodynamical relation in general context}
We start with a consideration that what relation we have between an entropy and energies when we adopt the arbitrary
form of entropy functional and the energy constraints with respect to a probability set. This 
problem was considered for the case of the canonical ensemble, that is, with one energy constraint to the entropy to be extremized
\cite{APlast1}. We deal with the generalized entropy $S(\{ p_i\})$ and the generalized expectation value of 
observables(generalized energy) $E^\sigma(\{ p_i\},\{ M_i^\sigma\})$, where $p_i$ denotes a probability of the microstate $i$ of a quantity $M_i^\sigma$, $(i=1,\cdots ,W)$. The superscript $\sigma$ labels a constraint number 
$(\sigma =1,\cdots ,N)$. We usually suppose $N < W$ since we treat a huge number of microstates $W$. The important thing is that the information we have at first is the $N$ mean values $E^\sigma$ and each $p_i$ is not {\it a priori} known. The probability $p_i$ should be given in terms of the Jaynes Maximum entropy principle instead. 
Extremization $S$ with respect to $p_i$ subject to $N$ generalized energy and the normalization condition of the 
probability leads to
\begin{equation}
\frac{\delta}{\delta p_i}\left( S-\sum_{\sigma=0}^N \beta_\sigma E^\sigma\right)=0
\end{equation}
where we have introduced the $N+1$ Lagrange multipliers $\beta_\sigma$ ,$(\sigma=0,\cdots ,N)$ and set 
$E^0=\sum_i^Wp_i=1$, i.e.,
\begin{equation}
\frac{\partial S}{\partial p_i}-\sum_{\sigma=0}^N\beta_\sigma\frac{\partial E^\sigma}{\partial p_i}
=0.
\end{equation}
Since the solution in equilibrium $p_i^*$ should be of the form $p_i^*=p_i^*(\beta_0(\beta_1,\ldots ,\beta_N),\beta_1,\ldots ,\beta_N)$ with the normalization of $p_i^*$, 
we have the partial derivatives of $S$ and $E^\sigma$ with respect to the $\mu$-th Lagrange parameter,
\begin{equation}
\frac{\partial S}{\partial \beta_\mu}=\sum_{i=1}^W\frac{\partial S}{\partial p_i^*}
\left( \frac{\partial p_i^*}{\partial \beta_\mu}+\frac{\partial p_i^*}{\partial \beta_0}
\frac{\partial \beta_0}{\partial \beta_\mu}\right),
\end{equation}
and
\begin{equation}
\frac{\partial E^\sigma}{\partial \beta_\mu}=\sum_{i=1}^W\frac{\partial E^\sigma}{\partial p_i^*}
\left( \frac{\partial p_i^*}{\partial \beta_\mu}+\frac{\partial p_i^*}{\partial \beta_0}
\frac{\partial \beta_0}{\partial \beta_\mu}\right)\label{eqn:dEdb},
\end{equation}    
respectively. After multiplying Eq(\ref{eqn:dEdb}) by $\beta_\sigma$ and summing over $\sigma$, one finds
\begin{eqnarray}
\sum_{\sigma=0}^N\beta_\sigma\frac{\partial E^\sigma}{\partial \beta_\mu} & = & \sum_{i=1}^W\sum_{\sigma=0}^N
\beta_\sigma\frac{\partial E^\sigma}{\partial p_i^*}
\left( \frac{\partial p_i^*}{\partial \beta_\mu}+\frac{\partial p_i^*}{\partial \beta_0}
\frac{\partial \beta_0}{\partial \beta_\mu}\right)\nonumber\\
& = & \sum_{i=1}^W\frac{\partial S}{\partial p_i^*}
\left( \frac{\partial p_i^*}{\partial \beta_\mu}+\frac{\partial p_i^*}{\partial \beta_0}
\frac{\partial \beta_0}{\partial \beta_\mu}\right)\nonumber\\
& = & \frac{\partial S}{\partial \beta_\mu}\label{eqn:Euler1},
\end{eqnarray}
which gives the generalized thermodynamical relation that connects the arbitrary form of the entropy and the 
mean values,
i.e.,
\begin{eqnarray}
\frac{\partial S}{\partial E^\sigma} & = &\sum_{\nu =0}^N\frac{\partial S}{\partial \beta_\nu}\frac{\partial \beta_\nu}
{\partial E^\sigma}=\sum_{\nu =0}^N\sum_{\mu =0}^N\beta_\mu \frac{\partial E^\mu}{\partial \beta_\nu}
\frac{\partial \beta_\nu}{\partial E^\sigma}\nonumber\\
& = & \beta_\sigma \quad (\sigma\ne 0)\label{eqn:recip1}.
\end{eqnarray}
$N=1$ corresponds to the canonical ensemble theory and gives the fundamental thermodynamical relation
\cite{APlast1},
\begin{equation}
\frac{\partial S}{\partial E}=\beta
\end{equation}
The above relation constitutes the basis of the justification that the Lagrange multiplier $\beta$ appearing in the equilibrium statistical mechanics based on the BGS entropy is identified with the inverse temperature in thermodynamics, however, we can stress that this thermodynamical relation is very robust in general context if only there is one energy constraint. Although we give no explicit form of the entropy $S(\{p_i^*\})$ in the present consideration, the condition of extremization with respect to the each Lagrange multiplies with fixing general energies is to be required, namely,
\begin{equation}
\frac{\partial S\{p_i^*\}}{\partial \beta_\mu}\Big|_{E^1,\cdots ,E^N}=0.
\end{equation}

\section{Legendre transform structure}
Consider now the following general entropic form
\begin{equation}
S=\sum_{i=1}^Wf(p_i)\label{eqn:Sf},
\end{equation}
where a measure $f(p_i)$ is an arbitrary function of the probability $\{p_i\}$ and we do not have to 
require the concavity property which is needed in physical stability in the following discussion. It should be 
noted that one form $f(p_i)=-p_i\ln p_i$ is the BGS measure when we choose Boltzmann's constant as the information unit and another form $f(p_i)=(p_i-p_i^q)/(q-1)$, $(q\in \mathbb{R})$ is Tsallis one. 
Moreover let us define the mean value of quantities (eigenvalues)$\{M_i^\sigma\}$ in the following way,
\begin{equation}
\langle M^\sigma \rangle =\sum_{j=1}^W g_j(p_1,\ldots ,p_W)M_j^\sigma \quad (\sigma = 1,\cdots N),
\end{equation}
where $g_j(p_1,\ldots ,p_W)$ determined by all probabilities is a weighting function for eigenvalues $M_j^\sigma$. 
In ordinary case, we should constrain $g_j$ as being $\sum_{j=1}^Wg_j=1$ from physically acceptable definition of the mean value (the mean value of unity should be unity). As a specific form of $g_j$, an escort type of probability\cite{Beck} satisfies this condition 
(hereafter we refer to as a weighting condition) as a weighting function,
\begin{equation}
g_j(p_1,\ldots ,p_W)=\frac{\phi (p_j)}{\sum_{i=1}^W\phi (p_i)}
\end{equation}
where $\phi (p_j)$ is a positive test function defined for all $p_j \in [0,1]$.
It is worth noting that the special test function of the form $\phi (p_j) =p_j^q$ constitute the 
generalized expectation value in Tsallis statistical mechanics\cite{98Tsallis}. In the present consideration, 
however, we weigh the eigenvalues with an general $g_j$ to proceed to a discussion. 
We will resultingly see that our conclusion can be derived even without the weighting condition. 
\\

The variational approach applied to extremizing Eq(\ref{eqn:Sf}) in this context is quite similar to 
the one developed in Sec.2
\begin{equation}
\frac{\delta}{\delta p_i}\left[ S-\alpha\sum_{i=1}^Wp_i -\sum_{\sigma =1}^N \beta_\sigma \sum_{j=1}^WM_j^\sigma 
g_j(p_1,\ldots,p_W)\right]=0
\end{equation}
where $\alpha$ and $\beta_\sigma$ ($\sigma =1,\cdots ,N$) are Lagrange multipliers again yielding to 
\begin{equation}
f'(p_i)-\alpha -\sum_{\sigma =1}^N \beta_\sigma \sum_{j=1}^W\frac{\partial g_j(p_1,\ldots ,p_W)}{\partial p_i}
M_j^\sigma =0.
\end{equation}
The {\it prime} denotes the derivative with respect to $p_i$. For a given set of eigenvalues 
$\{ M_j^\sigma\}$, we define $Q$ as a function of the probability set $\{ p_i\}$ and 
the Lagrange multipliers set $\{ \beta_\sigma\}$ as follows,
\begin{equation}
Q(\{ p_i\},\{ \beta_\sigma \})=\sum_{j=1}^W\sum_{\sigma =1}^N\beta_\sigma M_j^\sigma g_j(p_1,\ldots ,p_W)\
\label{eqn:Qdef}.
\end{equation}
Then the equation which determine the equilibrium probabilities $\{ p_i^* \}$ becomes 
\begin{equation}
P(p_i)=\alpha +\frac{\partial}{\partial p_i}Q(\{ p_i\},\{ \beta_\sigma \})\label{eqn:Pp},
\end{equation}
where we put $f'(p_i)=P(p_i)$.
Although the explicit form of the solution $\{p_i^*\}$ is not obtained 
from the above, we can proceed further discussion along the line of \cite{APlast1} 
by regarding the solution as $\{ p_i^*(\{ \beta_\sigma \}_{\sigma =1,\ldots ,N})\}$. 
Therefore, the $S$ and the $\langle M^\sigma\rangle$ read 
\begin{equation}
S=\sum_{i=1}^Wf(p_i^*)
\end{equation}
and
\begin{equation}
\langle M^\sigma \rangle = \sum_{j=1}^Wg_j(p_1^*,\ldots ,p_W^*) M_j^\sigma
\end{equation}
respectively. From this we immediately have 
\begin{equation}
\frac{\partial S}{\partial \beta_\sigma}=\sum_{i=1}^Wf'(p_i^*)\frac{\partial p_i^*}
{\partial \beta_\sigma}
\end{equation}
and
\begin{equation}
\frac{\partial \langle M^\mu \rangle}{\partial \beta_\sigma}
= \sum_{j=1}^W\sum_{i=1}^W\frac{\partial g_j(p_1^*,\ldots ,p_W^*)}
{\partial p_i^*}\frac{\partial p_i^*}{\partial \beta_\sigma}M_j^\mu \label{eqn:Mdbeta}.
\end{equation}

We are considering the Legendre transform of the entropy $S$
\begin{equation}
{\mathcal L} [S]=S-\sum_{\sigma =1}^N
\frac{\partial S}{\partial \langle M^\sigma\rangle}\langle M^\sigma\rangle.
\end{equation}
To evaluate the above, let us calculate the derivatives of the general entropy $S$ in equilibrium with 
respect to both $\beta_\sigma$ and $\langle M^\sigma\rangle$.
With Eq(\ref{eqn:Pp}) and Eq(\ref{eqn:Mdbeta}), $\partial S/\partial\beta_\sigma$ becomes
\begin{equation}
\frac{\partial S}{\partial \beta_\sigma} = \sum_{i=1}^W(\alpha +\frac{\partial Q}{\partial p_i^*})
\frac{\partial p_i^*}{\partial \beta_\sigma}=\sum_{i=1}^W\frac{\partial Q}{\partial p_i^*}
\frac{\partial p_i^*}{\partial \beta_\sigma}\label{eqn:dsdb}
\end{equation}
where we have used the relation arising from the normalization condition
\begin{equation}
\sum_{i=1}^W\frac{\partial p_i^*}{\partial \beta_\sigma}=\frac{\partial}{\partial \beta_\sigma}
\sum_{i=1}^Wp_i^*=0.
\end{equation}

Then from Eq(\ref{eqn:Qdef}), Eq(\ref{eqn:dsdb}) immediately leads to
\begin{eqnarray}
\frac{\partial S}{\partial \beta_\sigma} & = & \sum_{\mu =1}^N\beta_\mu \left[ \sum_{i=1}^W\sum_{j=1}^W 
M_j^\mu \frac{\partial g_j(p_1,\ldots ,p_W)}{\partial p_i^*}\frac{\partial p_i^*}
{\partial \beta_\sigma}\right]\nonumber\\ 
& = & \sum_{\mu =1}^N\beta_\mu \frac{\partial \langle M^\mu \rangle}
{\partial \beta_\sigma}\label{eqn:Euler2}. 
\end{eqnarray}
Therefore we have 
\begin{eqnarray}
\frac{\partial S}{\partial \langle M^\sigma\rangle} & = & \sum_{\nu =1}^N\frac
{\partial S}{\partial \beta_\nu}\frac{\partial \beta_\nu}
{\partial \langle M^\sigma\rangle}=\sum_{\nu =1}^N\sum_{\mu =1}^N\beta_\mu \frac{\partial \langle M^\mu \rangle}{\partial \beta_\nu}\frac{\partial \beta_\nu}
{\partial \langle M^\sigma\rangle}\nonumber\\
& = & \beta_\sigma \label{eqn:recip2}.
\end{eqnarray}
Thus, we again see that Eq(\ref{eqn:Euler2}) and Eq(\ref{eqn:recip2})  maintain the same relations as 
Eq(\ref{eqn:Euler1}) and Eq(\ref{eqn:recip1}) respectively as expected. Eq(\ref{eqn:Euler2})(Eq(\ref{eqn:Euler1})) 
expresses the Euler's Theorem in the general context\cite{APlast1,Aliaga}. Eq(\ref{eqn:recip2})
(Eq(\ref{eqn:recip1})) represents a general thermodynamical relation where the Lagrange multipliers and 
the mean values constitute conjugate variables each other with respect to $S$. It should be noted that this 
thermodynamical relation (namely, reciprocity relation\cite{APlast1}, which corresponds to the one relating the intensive parameters with the extensive parameters in ordinary thermodynamics) is seen to be independent of 
{\it both} the entropy functional and the mean energy form.

Since the entropy $S$ can be described either with entire set of $\langle M^\sigma\rangle$'s as  
$S(\{\langle M^\sigma\rangle\})$, or with $\beta_\sigma$'s as $S(\{\beta_\sigma\})$ due to the reciprocity relation, if we regard $S$ as $S(\{\langle M^\sigma\rangle\})$, the derivative of ${\mathcal L}[S]$ with respect to 
$\beta_\sigma$ becomes
\begin{eqnarray}
\frac{\partial {\mathcal L} [S]}{\partial \beta_\sigma} & = & \sum_{\nu =1}^N\frac{\partial S}{\partial \langle M^\nu\rangle}\frac{\partial \langle M^\nu\rangle}{\partial \beta_\sigma}-
\sum_{\nu =1}^N\beta_\nu\frac{\partial \langle M^\nu\rangle}{\partial \beta_\sigma} -\langle M^\sigma\rangle\nonumber\\
& = & -\langle M^\sigma\rangle\label{eqn:Lg1}
\end{eqnarray}
where we have used $\partial S/\partial \langle M^\nu \rangle=\beta_\nu$. In a similar way, 
if we regard the $S$ as $S(\{\beta_\sigma\})$, that is, 
\begin{equation}
{\mathcal L} [S] =S(\{\beta_\sigma\}) - \sum_{\sigma =1}^N\beta_\sigma \langle M^\sigma\rangle (\{\beta_\sigma\})
\end{equation}
then the derivative of ${\mathcal L}[S]$ with respect to
$\langle M^\sigma \rangle$ gives 
\begin{eqnarray}
\frac{\partial {\mathcal L} [S]}{\partial \langle M^\sigma\rangle} & = & \sum_{\nu =1}^N\frac{\partial S}
{\partial \beta_\nu}\frac{\partial \beta_\nu}{\partial \langle M^\sigma\rangle}-
\sum_{\nu =1}^N\frac{\partial \beta_\nu}{\partial \langle M^\sigma\rangle}\langle M^\nu\rangle-\beta_\sigma\nonumber\\
& = & -\sum_{\nu =1}^N\frac{\partial \beta_\nu}{\partial \langle M^\sigma\rangle}\langle M^\nu\rangle\nonumber\\
& = & -\sum_{\nu =1}^N\frac{\partial^2 S}{\partial \langle M^\sigma\rangle 
\partial \langle M^\nu\rangle}\langle M^\nu\rangle
\end{eqnarray}
where we have used 
\begin{equation}
\sum_{\nu =1}^N\sum_{\mu =1}^N\beta_\mu\frac{\partial \langle M^\mu\rangle}{\partial \beta_\nu}
\frac{\partial \beta_\nu}{\partial \langle M^\sigma\rangle}=\beta_\sigma
\end{equation}
in the first term of the first line.\\
Further, from Eq(\ref{eqn:recip2}),Eq(\ref{eqn:Lg1}) we have 
\begin{equation}
\frac{\partial \langle M^\sigma \rangle}{\partial \beta_\sigma}=
-\frac{\partial^2 {\mathcal L}[S]}{\partial \beta_\sigma^2}=\frac{1}{\frac{\partial^2 S}
{\partial \langle M^\sigma \rangle^2}}\label{eqn:Stab}.
\end{equation}
To guarantee the uniqueness of the LTS, that is, from a requisition that $\partial {\mathcal L}[S]/\partial \beta_\sigma$ and $\partial S/\partial \langle M^\sigma\rangle$ are to be monotonic functions, we 
assume 
\begin{equation}
\frac{\partial^2 {\mathcal L}[S]}{\partial \beta_\sigma^2}\ne 0,\quad 
\frac{\partial^2 S}{\partial \langle M^\sigma\rangle^2}\ne 0.
\end{equation}  
As indicated by Eq(\ref{eqn:Stab}), we see that when the $S$ is convex, then the ${\mathcal L}[S]$ is concave, and vice versa in general context. From the Eqs(\ref{eqn:recip2}),(\ref{eqn:Lg1}) and 
(\ref{eqn:Stab}), we can stress that the LTS is preserved {\it both} for an arbitrary form of 
the entropy and for the general definition of the mean value.

\section{Gibbs' fundamental equation}
We have found that the given mean values $\langle M^\sigma\rangle$ and the associated Lagrange 
multipliers $\beta_\sigma$ are 'thermally conjugated\cite{Beck}' each other in general context.
This relation (Eq(\ref{eqn:recip2})) for all $\sigma$ can be written in the following form
\begin{equation}
dS=\sum_{\sigma =1}^N \beta_\sigma d\langle M^\sigma\rangle\label{eqn:Gibbs},
\end{equation}
which is called Gibbs' fundamental equation\cite{Beck,Schlogl}. 
The concrete expression for Eq(\ref{eqn:Gibbs}) in the thermodynamics is well known as
\begin{equation}
dS=\frac{1}{T}dU+\frac{p}{T}dV-\frac{\mu}{T}dN.
\end{equation}
We should notice that, in the conventional phenomenological thermodynamic picture, the entropy $S$, 
the internal energy $U$, the volume $V$ and the particle number $N$ are regarded as extensive parameters and the coefficients ($T$:temperature, $p$:pressure, $\mu$:chemical potential) are intensive parameters, however, Gibbs' fundamental equation holds independently of the entropic form (extensive or nonextensive) and of the definition (linear or nonlinear) of the expectation value. 

\section{Summary and Conclusion}
We have derived the thermodynamical relation between the entropy and the energy in most general 
circumstance (Eq(\ref{eqn:recip2})) along the line of \cite{APlast1}. Furthermore, 
we have shown that the Legendre transform structure is a robust structure against 
the choice of the statistical entropic measure and the way of weighting for energy eigenstates of the system 
under consideration. As a necessary consequence of this structure, the Gibbs' fundamental 
equation is also independent of the present extension into the arbitrariness. 
These results are supported by the ubiquitous property of the Jaynes maximum entropy principle. The construction of a statistical mechanics whose basis should be consistent with the conventional thermodynamics, therefore, 
is considered to be not adequate with the preservation of the LTS and Gibbs' fundamental equation alone, which of course are essential ingredient for realization. The present conclusion  
is considered to be consistent with results of Ref\cite{Mendes} which took another route for 
discussing arbitrary thermostatistics.\\

The author acknowledges referees for useful remarks and comments and adjudicator for suggesting him how to 
revise this paper from its original.   

\end{document}